\documentclass[useAMS,usenatbib]{aastex61} 
\usepackage{epsfig}
\usepackage{lineno}
\usepackage{epsfig}  
\usepackage{amsmath} 
\usepackage{amssymb}
\usepackage{natbib} 
\usepackage{graphicx} 
\usepackage{endnotes}
\usepackage{color}
\newcommand{\xte}{{\it RXTE}}
\newcommand{\asat}{{\it AstroSat}}
\newcommand{\inte}{{\it INTEGRAL}}
 \shortauthors{Rawat et al. }
 \shorttitle{variability transition in GRS 1915+105}

\begin{document}

\title{Study of timing evolution from non-variable to structured large-amplitude variability transition in GRS 1915+105 using \asat{}}
\correspondingauthor{Divya Rawat}
\email{divyar@iitk.ac.in}

\author{Divya Rawat}
\affiliation{Department of physics, IIT Kanpur, Kanpur, Uttar Pradesh 208016, India}

\author{Mayukh Pahari}
\affiliation{Department of Astronomy and Astrophysics, Tata Institute of Fundamental Research,
Colaba, Mumbai 400005, India}
\affiliation{Royal Society-SERB Newton International Fellow, School of Physics and Astronomy, University of Southampton, Southampton, SO17 1BJ, UK}

\author{J S Yadav}
\affiliation{Department of Astronomy and Astrophysics, Tata Institute of Fundamental Research,
Colaba, Mumbai 400005, India}

\author{Pankaj Jain}
\affiliation{Department of physics, IIT Kanpur, Kanpur, Uttar Pradesh 208016, India}

\author{Ranjeev Misra}
\affiliation{Inter-University Center for Astronomy and Astrophysics, Ganeshkhind, Pune 411007, India}

\author{Kalyani Bagri}
\affiliation{ Department of physics, Pt. Ravishankar Shukla University, Raipur, Chhattisgarh 492010, India}

\author{Tilak Katoch}
\affiliation{Department of Astronomy and Astrophysics, Tata Institute of Fundamental Research,
Colaba, Mumbai 400005, India}

\author{P C Agrawal}
\affiliation{ UM-DAE Centre for Excellence in Basic Sciences, University of Mumbai,Kalina, Mumbai, Maharashtra 400098, India}

\author{R K Manchanda}
\affiliation{University of Mumbai, Kalina, Mumbai-400098, India}
\label{firstpage}

\begin{abstract}
In this work, we present a $\sim$90 ks continuous monitoring of the Galactic micro-quasar GRS 1915+105 with \asat{} when the source undergoes a major transition from a non-variable, $\chi$ class (similar to radio-quiet $\chi$ class) to a structured, large amplitude, periodic heartbeat state (similar to $\rho$ class). We show that such transition takes place via an intermediate state when the large-amplitude, irregular variability of the order of hundreds of seconds in the soft X-ray band turned into a 100-150 sec regular, structured, nearly periodic flares. The properties of a strong low-frequency quasi-periodic oscillations (LF QPO) in the frequency range 3-5 Hz also evolve marginally during these variability transitions. We also study time-lag and rms spectra at the QPO and harmonic component and the dynamic power spectra. We note few important differences between the heartbeat state and the $\rho$ class. Interestingly, the time-averaged LF QPO properties in the hard X-ray band is relatively stable in three states  when compared to the significant evolution observed in the slow variability properties at mHz frequencies. Such relative stability of LF QPOs implies the inner disk-corona coupled accretion flow which determines the LF QPO properties, may be uninterrupted by the launch of long, large-amplitude flares.  
\end{abstract}


\section{Introduction}\label{intro}
Due to its extremely rich and puzzling variability features, GRS 1915+105 is the most studied Galactic micro-quasar covering from the radio to $\gamma$-ray band. Among all bands, it is best-studied in X-rays since the X-ray luminosity changes from few percents of Eddington luminosity to the near-Eddington luminosity in few tens of seconds. Such extreme activities in this source  have continued for last 25 years since it was discovered with WATCH instrument onboard the Granat satellite in 1992. Depending upon the temporal variability pattern, hardness ratio, and color-color diagram, its lightcurve can be categorized into 14 different classes \citep{be00,ha05}. All the classes can be explained in terms of 3 distinct spectral states A, B, and C such that all the different classes can be regarded as a transition between these three states. Among 14 classes, $\chi$ and $\phi$ are the least variable classes, i.e., no long-term, large-amplitude variability in X-ray count rate is observed from both classes \citep{be00,tr01,mu01}. While the $\chi$ class is spectrally hard, the $\phi$ class is spectrally soft. On the other hand, among large-amplitude variability classes, $\rho$ class is most frequently observed and most-studied because of it's structured, regular and highly periodic X-ray lightcurve profile \citep{pa98, ya99}, also called heartbeat oscillations \citep{ne11,ne12}. The large-amplitude variability $\kappa$ and $\lambda$ classes are not observed as often \citep{be97a,be97b}. The origin of such large amplitude variability is usually associated with the radiation-pressure instability due to its persistently high, near-Eddington mass accretion rate \citep{ja00}. In another approach, \citet{na01} showed that a modified viscosity law could reproduce large-amplitude variability qualitatively. Although characteristics and the nature of individual class have been studied well in long observations \citep{ya99,gr96,na00}, the gradual transition from one class to another, that requires long and continuous observations have rarely been studied.  

Using 2-years of X-rays (\inte{} and \xte{}) and radio (Ryle) monitoring of GRS 1915+105, \citet{ro08a} found class transitions and show that discrete radio emission is evident in GRS 1915+105. This arises as a response to an X-ray succession consisting of a spectrally hard X-ray dip terminated by a narrow X-ray spike marking the disappearance of the hard Comptonized X-ray emission. These have been earlier studied by \citet{mi98,ei98}. They identify X-ray spike as the trigger of the radio ejection and a correlation between the flux density of radio flares and the duration of the X-ray dips. Using spectro-temporal analysis in a subsequent paper, \citet{ro08b} showed that X-ray spikes marked coronal mass ejections while the powerlaw index correlates with the radio flux density. They found the relative contribution of Comptonized flux in comparison to the total flux changes in the spectrum of quasi-periodic oscillations and therefore rule out the global-mode coronal oscillation model as the origin of QPOs.  

While studying the long-term evolution of the $\omega$ class in GRS 1915+105, defined by a sequence of short X-ray dips at nearly regular intervals, \citet{pa10} found that with the increasing mass accretion rate, the X-ray dips become infrequent and gradually disappear, leaving behind a high soft, non-variable continuum in the lightcurve. Among different classes in GRS 1915+105, a major focus is given to understanding the physics of $\rho$ class or heartbeat oscillation. In the color-color diagram, pulse profiles create a loop-like pattern that rotates clockwise \citep{be00}. The $\rho$ class sometimes show a single-peaked soft pulse and sometimes a double pulse combining a sequence of soft and hard pulses. Using detailed phase-resolved spectroscopy with Chandra data \citet{ne11,ne12} showed that spectral and timing properties of the heartbeat like X-ray pulses are in agreement with the combined effect of the radiation pressure instability and a local Eddington limited radiation mechanism in the accretion disk and radiation-driven wind. During the hard pulse, they noted a burst of Bremsstrahlung radiation. The $\rho$ class has also been observed from another black hole X-ray transient IGR J17091-3624 \citep{al11,pa14,co17}. 

Large Area X-ray Proportional Counter (LAXPC) on-board \asat{} observed the radio-quiet $\chi$ class in GRS 1915+105 and studied the detailed timing properties \citep{ya16b}. Not only in GRS 1915+105, detail spectro-timing studies have been performed using LAXPC observations in other X-ray binaries like Cyg X-1 \citep{mi17},  4U 1728-34 \citep{ve17} and Cyg X-3 \citep{pa17, pa18}. In this work, using \asat{} observations we present continuous monitoring of GRS 1915+105 when a gradual transition from a non-variable state (similar to $\chi$ class) to the highly structured, large-amplitude heartbeat state (similar to $\rho$ class) via an intermediate state of irregular flaring, prominently visible in the soft X-ray band. Comparing timing properties with previous studies, we note that the $\chi$ class as observed by the AstroSat/LAXPC has few dissimilarities with that observed by the RXTE/PCA \citep{pa13}. Such dissimilarities can be attributed to the higher detection efficiency of the LAXPC compared to the PCA at the hard X-ray bands. The heartbeat state as observed by LAXPC, has significant morphological differences with typical $\rho$ class as observed by RXTE/PCA \citep{ne12}. Interestingly, a low-frequency QPO is observed between 3-5 Hz in all observations which shows similar properties during the non-variable and heartbeat state. These observations have interesting physical implications which we describe in the discussion
section of this paper. 

\section{Observation and  Data Analysis}
\begin{figure*}
\centering\includegraphics[width=0.65\textwidth,angle=-90]{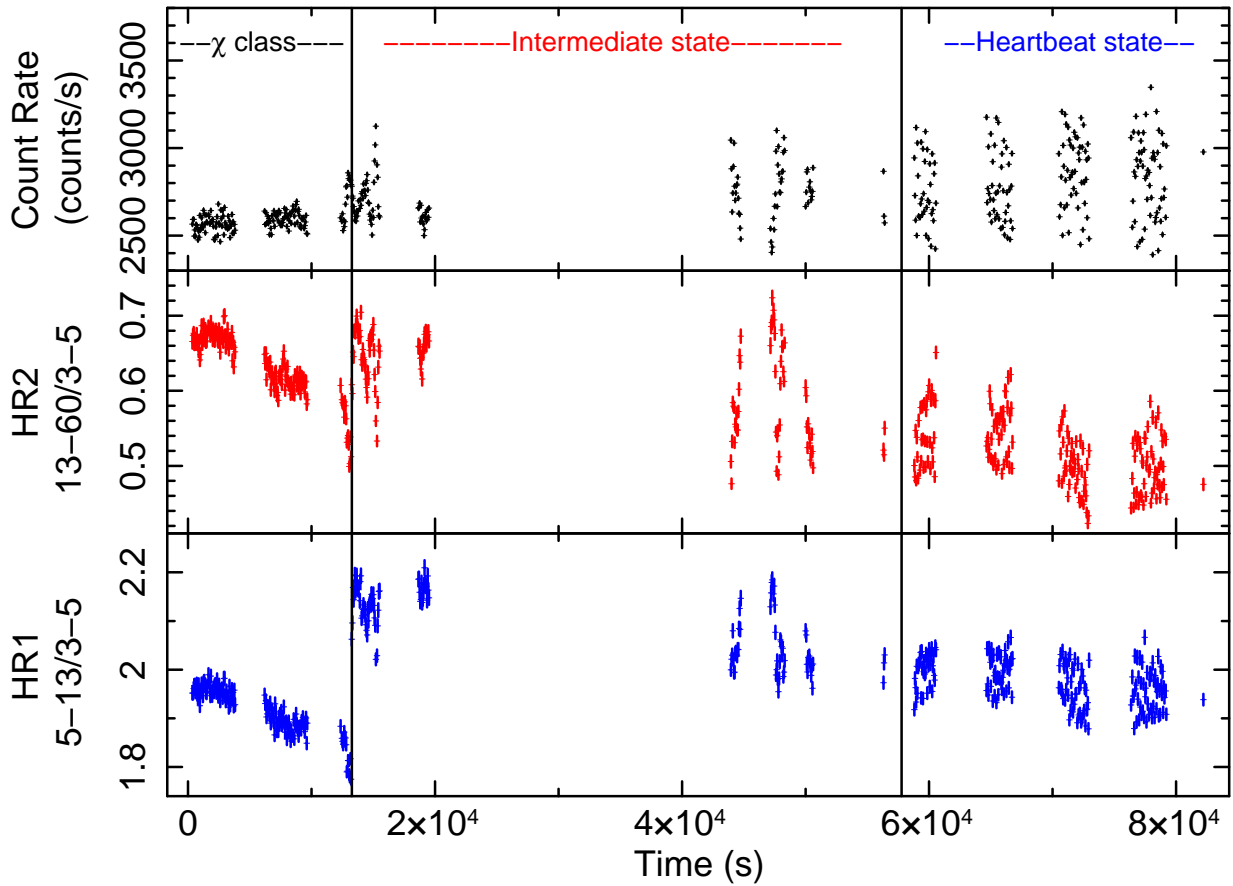}
\caption{A 50 sec binned complete light curve of GRS 1915+105 in 3.0-60.0 keV energy range (top panel) combining {\tt LAXPC10}, {\tt LAXPC20} and {\tt LAXPC30},  hard color (middle panel) and soft color is shown as a function of time.}
\label{lightcurve}
\end{figure*}

\begin{figure*}
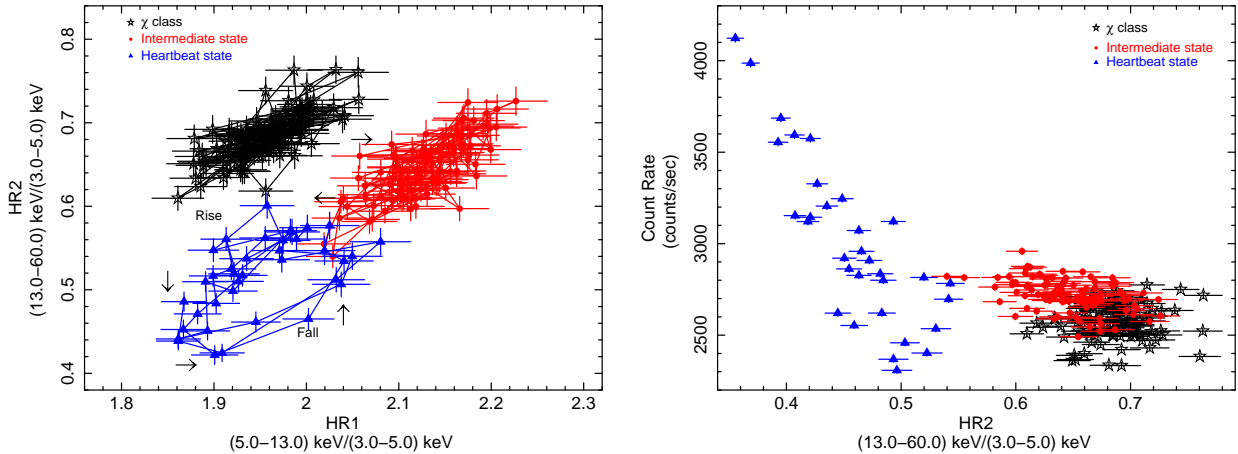

\centering \includegraphics[width=0.33\textwidth,angle=-90]{figure2.1.ps}
\centering \includegraphics[width=0.33\textwidth,angle=-90]{figure2.2.ps}
\caption{ Left panel shows the color-color diagram during three states where the hard color (HR2) is plotted against the soft color (HR1). We have used  first $\sim$1000 secs lightcurve of both $\chi$ class and IMS state and  last $\sim$500 secs lightcurve of HS for producing color-color diagram. Right panel shows hardness-intensity diagram where X-ray intensity in 3-60 keV is plotted against the hard color combining all LAXPC units. A 10 sec binned lightcurve is used for producing the plots.} 
\label{full_lc}
\end{figure*}

As a part of the Guaranteed Time observation, GRS 1915+105 is observed for $\sim$90 ks using Large Area X-ray Proportional Counter (LAXPC) instrument onboard \asat{} on 28 March 2017.  LAXPC consists of three identical but independent X-ray proportional counters which can register events with the absolute time resolution of 10 $\mu$s in the energy range 3.0-80.0 keV \citep{ya16a, ya16b, ag17}. LAXPC data are analyzed using the Laxpc software\footnote{\label{note2}\url{http://www.tifr.res.in/~astrosat_laxpc/LaxpcSoft.html}}. Details of the response and background spectra generation can be found in \citet{an17}. In Figure \ref{lightcurve} a 50.0 sec binned 3.0-60.0 keV full lightcurve, HR2 versus Time and HR1 versus Time plots are shown. The gaps in the light curve are due to the South Atlantic Anomaly (SAA), data loss and the Earth occultations. The hard color (HR2) is defined as the ratio of X-ray count rate in 13-60 keV and 3-5 keV while the soft color (HR1) is defined as the ratio of X-ray count rate in 5-13 keV and 3-5 keV. The color-color diagram (CCD) is defined as the plot of hard color versus soft color while the hardness-intensity diagram (HID) is defined as the plot of 3-60 keV X-ray intensity as a function of hard color. Based on the CCD and the HID plots as shown in the left and right panels of Figure \ref{full_lc}, we classify our observation into three states (i) Non-variable $\chi$ class, which is quite a stable state with no long-term variability in both soft and hard bands and occupies the hardest region in the CCD and HID. (ii) Heartbeat state (HS), which shows regular, periodic and structured flares in its lightcurve that resemble the $\rho$ class variability in GRS 1915+105. The hard color value is lowest during HS and it produces a loop in the CCD. We have used 10 sec binned light curve for generating CCD of HS so that the loop in CCD can be easily visualized (the time period of oscillation is $\sim$150 sec). (iii) Intermediate state (IMS), which is a state between the $\chi$ class and HS that shows large-amplitude, long timescale irregular variability  between 5-8 $\%$ (Table \ref{obs1}).

To check the presence of Quasi-periodic oscillations (QPO) in  the light curve we extract the power density spectrum (PDS) for the three states $\chi$ class, IMS, and HS. The lightcurve is segmented into 24 segments where each segment is binned at 8.39 ms and consists of 16384 points. Fourier transform of an individual segment is calculated and averaged over all segments. Rebinning of the resultant power spectrum in frequency space is done. All PDS are dead time-corrected (assuming a dead time of 42 $\mu$s) and Poisson noise has been estimated and subtracted (for details see \citet{ya16b, mi17}). 
The PDS are normalized such that
its integral over frequency equals the fractional rms squared.
Furthermore, the PDS are corrected for the corresponding background rates
\citep{ya16b}.  

To study the behavior of the time series as a function of energy we determine the time lag as a function of photon energy. We also
determine the  fractional rms as a function of energy for the QPO and the harmonic component for $\chi$ class and IMS respectively. As we shall 
see the harmonic component is not present in HS. Hence, in this case, we determine the time lag as a function of energy for the LF QPO 
and mHz QPO. Eight energy bands 3.0-4.0, 4.0-5.0, 5.0-7.0, 7.0-9.0, 9.0-12.0, 12.0-15.0, 15.0-20.0 and 20.0-30.0 keV are used for calculating the time-lag. Here 3.0-4.0 keV band is the reference energy band and the Fourier frequency is binned at an interval of 0.78 Hz and 1.25 mHz for LF QPO and mHz QPO respectively. Energy-dependent time lag and rms among different energy bands are computed following \citet{no99}.

We also determine the time-lag between 3-5 keV and 9-12 keV photons as a function of Fourier frequency using the LAXPC10 and LAXPC20 
 separately for the $\chi$ class, IMS and the HS. We use
both LAXPC10 and LAXPC20 data independently in order to check the consistency between
the two detectors. The Fourier frequency is binned at an interval of 0.6 Hz and 2.5 mHz for (1-16) Hz and (0.003-0.02)Hz frequency range respectively.

\section{Results}

\begin{figure*}
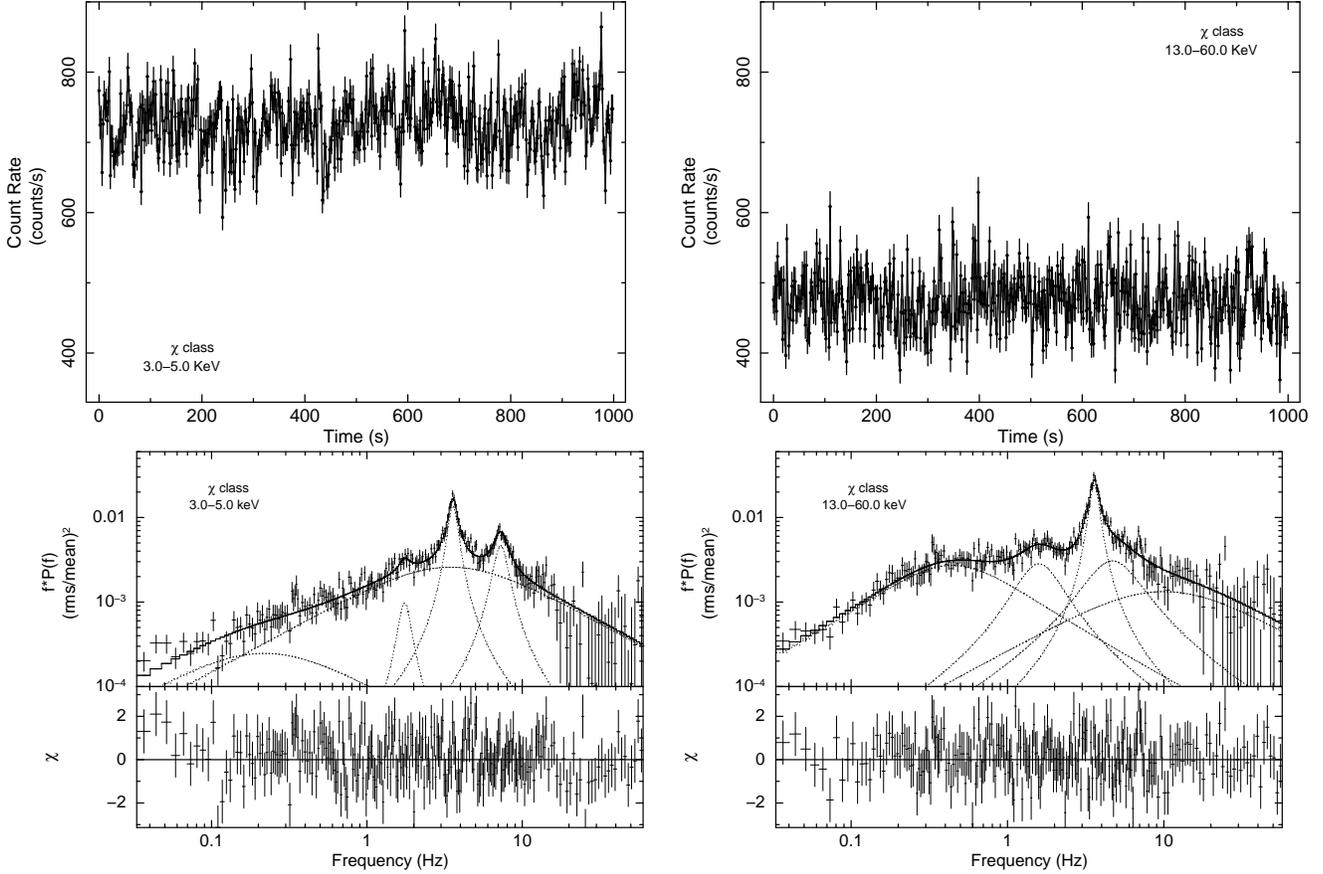

\centering \includegraphics[width=0.33\textwidth,angle=-90]{figure3.1.ps}
\centering \includegraphics[width=0.33\textwidth,angle=-90]{figure3.2.ps}
\centering \includegraphics[width=0.31\textwidth,angle=-90]{figure3.3.ps}
\centering \includegraphics[width=0.31\textwidth,angle=-90]{figure3.4.ps}
\caption{A 1000 sec lightcurve for $\chi$  class is shown in 3.0-5.0 and 13.0-60.0 keV energy range in top left and top right panels respectively using a bin time of 2 sec combining all LAXPC units. Bottom panel shows the back ground-corrected, Poisson noise subtracted power density spectra for $\chi$ class in 3.0-5.0 keV  and 13.0-60.0 keV energy range fitted with multiple Lorentzians.}
\label{Non}
\end{figure*}

\begin{figure*}
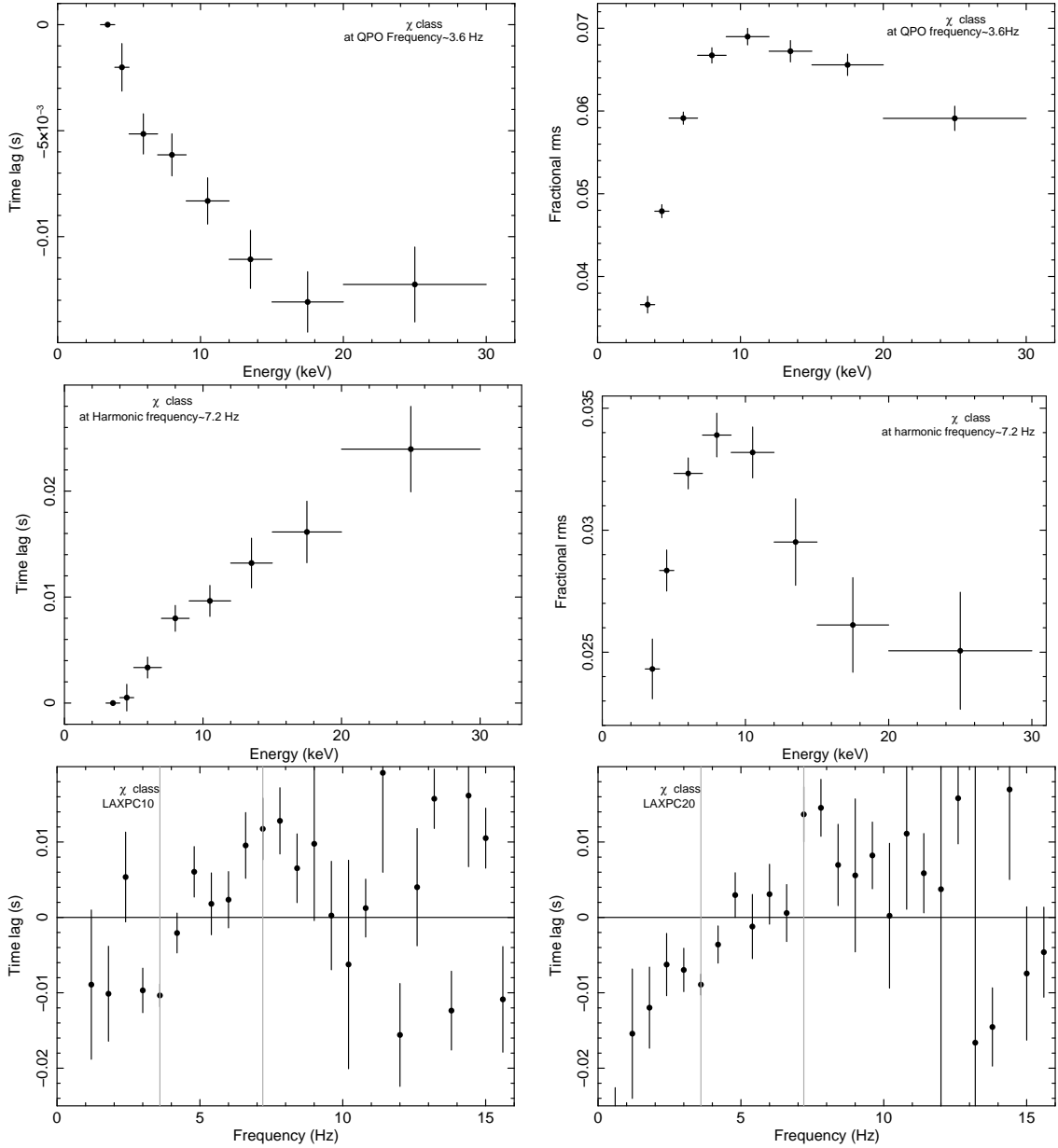

\centering \includegraphics[width=0.30\textwidth,angle=-90]{figure4.1.ps}
\centering \includegraphics[width=0.30\textwidth,angle=-90]{figure4.2.ps}
\centering \includegraphics[width=0.30\textwidth,angle=-90]{figure4.3.ps}
\centering \includegraphics[width=0.30\textwidth,angle=-90]{figure4.4.ps}
\centering \includegraphics[width=0.30\textwidth,angle=-90]{figure4.5.ps}
\centering \includegraphics[width=0.30\textwidth,angle=-90]{figure4.6.ps}
\caption{Time lag as a function of photon energy are shown at the QPO frequency$\sim$3.6 Hz (top left) and the harmonic frequency $\sim$7.2 Hz (middle left) during the $\chi$ class combining all LAXPC units. Fractional rms as a function of photon energy is shown at the QPO frequency$\sim$3.6 Hz (top right), the harmonic frequency $\sim$7.2 Hz (middle right) combining all LAXPC units. Time lag between 3-5 keV and 9-12 keV as a function of Fourier frequency is shown in the bottom left and bottom right panels for LAXPC10 and LAXPC20 respectively. Vertical grey lines in both panels show the position of the QPO and its harmonic.}
\label{timing1}
\end{figure*}
In Table \ref{obs1} we show various parameters for the three states orbitwise. It
shows that the Lightcurve fractional rms is increasing steadily from $\chi$ class to HS. The left panel of Figure \ref{full_lc} reveals some very important
details about the evolution of source from $\chi$ class to HS. When source
evolves from $\chi$ class to IMS, accretion disk temperature increases as
indicated by the increase in HR1. However we see that HR2 is remaining
practically unchanged during this transition. In its evolution from IMS
to HS, we find that the source moves along the diagonal spread (of data
points) towards lower values of HR1 and HR2. For the IMS, data points in
CCD plot occupy a position with a higher value of HR1 as
compared with $\chi$ class and HS. From the right panel of Figure \ref{full_lc}, an overall
decline in the hard color value is observed with
time which emphasizes the fact that source is making a transition from a
harder to a softer state.  Below we explore the properties of the individual states.

\subsection{$\chi$ class}
The top left and right panels of  Figure \ref{Non} show a 1000 secs light curve of $\chi$ class in 3.0-5.0 keV and 13.0-60.0 keV energy range respectively. We may see that X-ray flux level is smaller by a factor of 1.5 in hard energy band as compared to soft energy band. A short time variability of $<$10 secs is seen but no long-term variability ($\sim$100 secs) is observed in the light curve of this class. The PDS for $\chi$ class fitted using Lorentzians in soft (3.0-5.0) and hard (13.0-60.0) energy bands are shown in the bottom left and the bottom right panel of Figure \ref{Non}. A QPO and harmonic component at $\sim$3.6 Hz and $\sim$7.2 Hz is observed in the PDS of $\chi$ class for the soft band only. The harmonic component is not significant in 13.0-60.0 keV band. The QPO parameters for all the states and three energy bands (3.0-5.0, 5.0-13.0 and 13.0-60.0 keV) are given in Table \ref{obs2}, \ref{obs3} and \ref{obs4}. Two broad Lorentzian noise (BLN) components and  one sub-harmonic component are also observed in PDS.  As shown in Table \ref{obs2}, \ref{obs3} and \ref{obs4} a higher value of QPO fractional rms in 5.0-13.0 keV and 13.0-60.0 keV  as compared to 3.0-5.0 keV is noticed for all three states. 

In the top left panel of Figure \ref{timing1}, a decrease in the time lag at the higher energy band implies soft lag at the QPO frequency of $\sim$3.6 Hz. On the other hand, at the harmonic frequency of $\sim$7.2 Hz, a characteristic hard lag is observed. Although the reason is not clear, the time-lag at a given energy band is higher at the harmonic frequency compared to that of the QPO frequency. This implies that the phase lag between two given energy bands is different at the QPO and harmonic frequencies.

\begin{table*}
 \centering
 \caption{LAXPC observation details of GRS 1915+105}
\begin{center}
\scalebox{1.0}{
\begin{tabular}{ccccc}
\hline  
Orbit number & State &  Effective & Average count  rate & Lightcurve   \\  
& & exposure (s) & (3.0-80.0 keV)  &  fractional rms (\%)  \\ 
\hline
8107 & $\chi$ class & 3793 & $2562 \pm 0.9$ &  $3.3 \pm 0.1$ \\
8108 & $\chi$ class & 4059 & $2572  \pm 0.9$ & $3.2 \pm 0.1$  \\ 
8109 & $\chi$ class & 4569 & $2619 \pm 0.9$ & $3.7 \pm 0.1$ \\ 
8110 & IMS & 4325 & $2689 \pm 0.9$ & $6.1 \pm 0.2$  \\ 
8115 & IMS & 2971 & $2768 \pm 1.1$ & $6.8 \pm 0.3$ \\
8118 & HS & 2053 & $2769 \pm 1.3$ & $10.0 \pm 0.5$  \\
8119 & HS & 2198 & $2776 \pm 1.2$ & $9.0  \pm 0.4$  \\ 
8120 & HS & 2621 & $2878 \pm 1.2$ & $11.7 \pm 0.5$  \\ 
8121 & HS & 3195 & $2836 \pm 1.1$ & $13.8 \pm 0.6$  \\
\hline
\end{tabular}}
\tablecomments {IMS and HS stand for intermediate and Heartbeat state respectively. Average count rate is calculated combining all three LAXPC units. Lightcurve fractional rms is calculated using lcstats binned at an interval of 10.0 sec combining all LAXPC units.  }
\end{center}
\label{obs1}
\end{table*}

\begin{table*}
 \centering
 \caption{QPO Parameters for GRS 1915+105 in 3.0-5.0 keV enegy range}
\begin{center}
\scalebox{0.85}{%
\hspace{-2cm}
\begin{tabular}{ccccccccc}
\hline  
Orbit number & state & QPO frequency & QPO width & QPO fractional& Harmonic & width &fractional rms  & (0.0006-20) Hz \\  
& &  (Hz) &(Hz)& rms (\%) & frequency(Hz) & (Hz)& (\%) & total rms \%\\ 
\hline
8107 & $\chi$ class & $3.57 \pm 0.01$ & $0.56 \pm 0.04$ & $5.7 \pm 0.1$ & $7.21  \pm 0.06$& $1.66 \pm 0.20$& $4.0 \pm 0.3$ & $21.1 \pm 0.6$\\
8108 & $\chi$ class & $3.48\pm 0.01$ & $0.59 \pm 0.03$ & $5.7 \pm 0.1$ & $7.02  \pm 0.06$& $1.53 \pm 0.23$& $3.6 \pm 0.3$ &  $20.5 \pm 0.5$\\ 
8109 & $\chi$ class & $3.77 \pm 0.01$ &$0.63 \pm 0.04$ & $5.9 \pm 0.2$ &$7.67  \pm 0.05$& $1.65 \pm 0.20$& $3.7 \pm 0.3$  &  $21.3 \pm 0.7$\\ 
8110 & IMS & $4.16 \pm 0.01$ & $0.86 \pm 0.05$ & $5.8 \pm 0.2$ &$8.28  \pm 0.13$& $3.11 \pm 0.51$& $3.6 \pm 0.3$ &  $23.8 \pm 1.5$\\ 
8115 & IMS & $4.95 \pm 0.04$ & $1.22 \pm 0.11$ & $5.1 \pm 0.2$ &$9.96  \pm 0.31$& $5.83 \pm 1.68$& $4.0 \pm 0.2$ &  $23.2 \pm 2.3$\\
8118 & HS & $4.97 \pm 0.05$ & $1.26 \pm 0.14$ & $4.3 \pm 0.1$ & $10.13 \pm 0.31$& $5.14 \pm 1.60$& $3.9 \pm 0.4$ &  $24.0 \pm 2.5$\\
8119 & HS & $5.05 \pm 0.04$ & $1.18 \pm 0.13$ & $4.5 \pm 0.2$ & $10.27 \pm 0.47$& $5.47 \pm 1.49$& $3.4 \pm 0.4$ &  $24.0 \pm 3.6$\\ 
8120 & HS & $5.40 \pm 0.04$ & $1.21 \pm 0.11$ &$4.2 \pm 0.1$ & $10.36 \pm 0.75$& $10.33 \pm 2.73$& $3.8 \pm 0.4$ &  $33.4 \pm 2.3$\\ 
8121 & HS & $5.32 \pm 0.05$ & $1.51 \pm 0.15$ & $4.3 \pm 0.2$ & $9.82 \pm 0.63$& $8.97 \pm 2.25$& $4.10 \pm 0.4$ &  $36.4 \pm 2.9$\\ 
\hline
\end{tabular}}
\end{center}
\label{obs2}
\end{table*}

\begin{table*}
 \centering
 \caption{QPO Parameters for GRS 1915+105 in 5.0-13.0 keV energy range}
\begin{center}
\scalebox{0.85}{%
\hspace{-2cm}
\begin{tabular}{ccccccccc}
\hline  
Orbit number & state & QPO frequency & QPO width & QPO fractional & Harmonic  & width & fractional rms &(0.0006-20) Hz \\  
&   & (Hz) & (Hz) & rms  (\%) & frequency(Hz) & (Hz) &(\%)& total rms\%\\ 
\hline
8107 & $\chi$ class &$3.60 \pm 0.01$ &$0.63 \pm 0.03$ &$8.6 \pm 0.1$ &$7.07  \pm 0.04$ &$1.55  \pm 0.12$ &$4.1 \pm 0.1$ &$28.0 \pm 0.5$\\ 
8108 & $\chi$ class &$3.51 \pm 0.01$ &$0.65 \pm 0.03$ &$8.7 \pm 0.1$ &$6.90  \pm 0.03$ &$1.12  \pm 0.11$ &$3.7 \pm 0.1$ &$28.0 \pm 1.3$\\ 
8109 & $\chi$ class &$3.81 \pm 0.01$ &$0.70 \pm 0.03$ &$8.8 \pm 0.1$ &$7.48  \pm 0.05$ &$1.79  \pm 0.19$ &$3.9 \pm 0.1$ &$27.2 \pm 0.7$\\ 
8110 & IMS         &$4.21 \pm 0.01$ &$0.93 \pm 0.04$ &$8.5 \pm 0.2$ &$8.44  \pm 0.06$ &$2.27  \pm 0.19$ &$3.7 \pm 0.1$ & $32.7 \pm 1.9$\\ 
8115 & IMS         &$4.87 \pm 0.02$ &$1.32 \pm 0.07$ &$7.8 \pm 0.2$ &$9.51  \pm 0.16$ &$4.92  \pm 0.72$ &$4.3 \pm 0.5$ & $30.0 \pm 1.8$\\
8118 & HS   &$4.99 \pm 0.03$ &$1.48 \pm 0.08$ &$7.5 \pm 0.2$ &$8.96  \pm 0.46$ &$10.30 \pm 0.77$ &$6.7 \pm 0.2$ & $29.9 \pm 3.1$\\
8119 & HS  &$5.05 \pm 0.02$ &$1.33 \pm 0.08$ &$7.5 \pm 0.1$ &$9.07  \pm 0.47$ &$10.87 \pm 0.85$ &$6.6 \pm 0.2$ & $34.7 \pm 2.3$\\ 
8120 & HS   &$5.47 \pm 0.02$ &$1.42 \pm 0.06$ &$7.0 \pm 0.1$ &$10.72 \pm 0.30$ &$9.69  \pm 0.84$ &$5.4 \pm 0.1$ & $34.5 \pm 2.3$\\ 
8121 & HS   &$5.35 \pm 0.02$ &$1.41 \pm 0.05$ &$7.0 \pm 0.1$ &$8.96  \pm 0.46$ &$9.85  \pm 0.73$ &$5.8 \pm 0.1$ & $40.1 \pm 2.9$\\ 
\hline
\end{tabular}}
\end{center}
\label{obs3}
\end{table*}

\begin{table*}
 \centering
 \caption{QPO Parameters for GRS 1915+105 in 13.0-60.0 keV energy range}
\begin{center}
\scalebox{0.85}{%
\hspace{-2cm}
\begin{tabular}{ccccccccc}
\hline  
Orbit number & state & QPO frequency & QPO width & QPO fractional& Harmonic & width &  fractional rms & (0.0006-20) Hz \\  
& & (Hz) & (Hz) &rms  (\%)& frequency(Hz) &  (Hz) &(\%) & total rms \% \\ 
\hline
8107 & $\chi$ class & $3.60 \pm 0.01$ & $0.56 \pm 0.05$ & $7.6 \pm 0.4$ & $-$ & $-$ & $-$ & $29.2 \pm 1.6$\\ 
8108 & $\chi$ class & $3.48 \pm 0.01$ & $0.58 \pm 0.05$ & $8.1 \pm 0.4$ & $-$ & $-$ & $-$ &  $29.9 \pm 2.1$\\ 
8109 & $\chi$ class & $3.75 \pm 0.02$ & $0.55 \pm 0.06$ & $6.9 \pm 0.5$ & $-$ & $-$ & $-$ &  $28.1 \pm 2.3$\\ 
8110 & IMS         & $4.23 \pm 0.02$ & $0.99 \pm 0.08$ & $8.2 \pm 0.2$ & $-$ & $-$ & $-$ &  $32.6 \pm 2.5$\\ 
8115 & IMS         & $4.92 \pm 0.03$ & $0.89 \pm 0.16$ & $6.0 \pm 0.5$ & $-$ & $-$ & $-$ &  $28.6 \pm 3.7$\\
8118 & HS          & $4.98 \pm 0.05$ & $1.23 \pm 0.15$ & $4.2 \pm 0.2$ & $-$ & $-$ & $-$ & $26.9 \pm 1.7$\\
8119 & HS         & $5.01 \pm 0.05$ & $1.20 \pm 0.14$ & $6.8 \pm 0.5$ & $-$ & $-$ & $-$ &  $29.7 \pm 2.0$\\ 
8120 & HS         & $5.49 \pm 0.04$ & $1.68 \pm 0.12$ & $7.5 \pm 0.2$ & $-$ & $-$ & $-$ &  $30.2 \pm 2.2$\\ 
8121 & HS          & $5.31 \pm 0.03$ & $1.45 \pm 0.12$ & $7.3 \pm 0.3$ & $-$ & $-$ & $-$ &  $30.2 \pm 2.5$\\ 
\hline 
\end{tabular}}
\end{center}
\label{obs4}
\end{table*}

The strength of the QPO and harmonic component also varies with photon energy as shown in  the top right and middle right panels respectively in Figure \ref{timing1}. For QPO, the fractional rms rises to $\sim$7 percent and attains its maximum value at 9.0-12.0 keV energy band and decreases later to 5.5 percent. For the harmonic, the rms value reaches its maxima of $\sim$3.5 percent in 9.0-12.0 keV band and reduces significantly at higher energies. The reduction in the rms value of harmonic above 12.0 keV is the result of the absence of a harmonic component in hard energy band (13.0-60.0 keV). In the bottom panels of Figure \ref{timing1}, we see that the time-lag increases from soft lag at the lower frequency and makes a transition to the hard lag at a frequency of $\sim$5.5 Hz. Two dotted lines are marked at QPO and the harmonic frequency highlight the facts that time lag is negative for QPO and becomes positive for the harmonic component which is in agreement with the time-lag spectra shown in top panels of Figure \ref{timing1}.

\subsection{Intermediate state}

\begin{figure*}
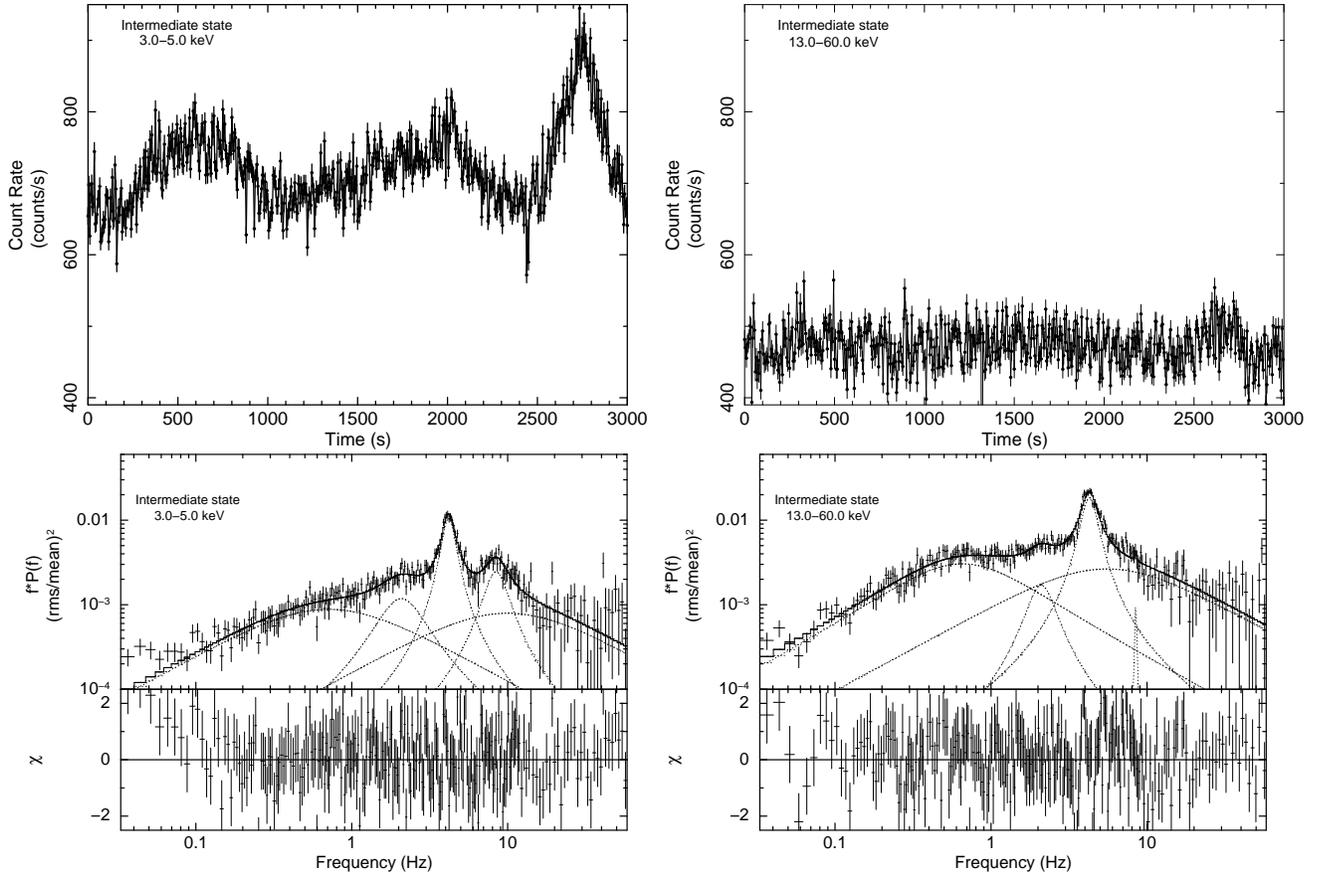

\centering \includegraphics[width=0.33\textwidth,angle=-90]{figure5.1.ps}
\centering \includegraphics[width=0.33\textwidth,angle=-90]{figure5.2.ps}
\centering \includegraphics[width=0.31\textwidth,angle=-90]{figure5.3.ps}
\centering \includegraphics[width=0.31\textwidth,angle=-90]{figure5.4.ps}
\caption{A $\sim$3000 sec section of the IMS lightcurve is shown in 3.0-5.0 and 13.0-60.0 keV energy range in top left and top right panels respectively using a bin time of 2 sec combining all LAXPC units. Bottom panels show the power density spectra for the IMS in 3.0-5.0 keV and 13.0-60.0 keV  energy range fitted with multiple Lorentzians.}
\label{intermediate}
\end{figure*}
The top left and right panels of Figure \ref{intermediate} show 3 ks lightcurve of the IMS, with a 2.0 sec bin size in two different energy ranges 3.0-5.0 keV and 13.0-60.0 keV respectively.  To account for the variability in the IMS, we extract PDS in 3.0-5.0 keV and 13.0-60.0 keV energy range. These are shown in the bottom left and bottom right panels of Figure \ref{intermediate} respectively. We fit the PDS with five Lorentzians: the QPO, harmonic, sub-harmonic and two broad noise components.

\begin{figure*}
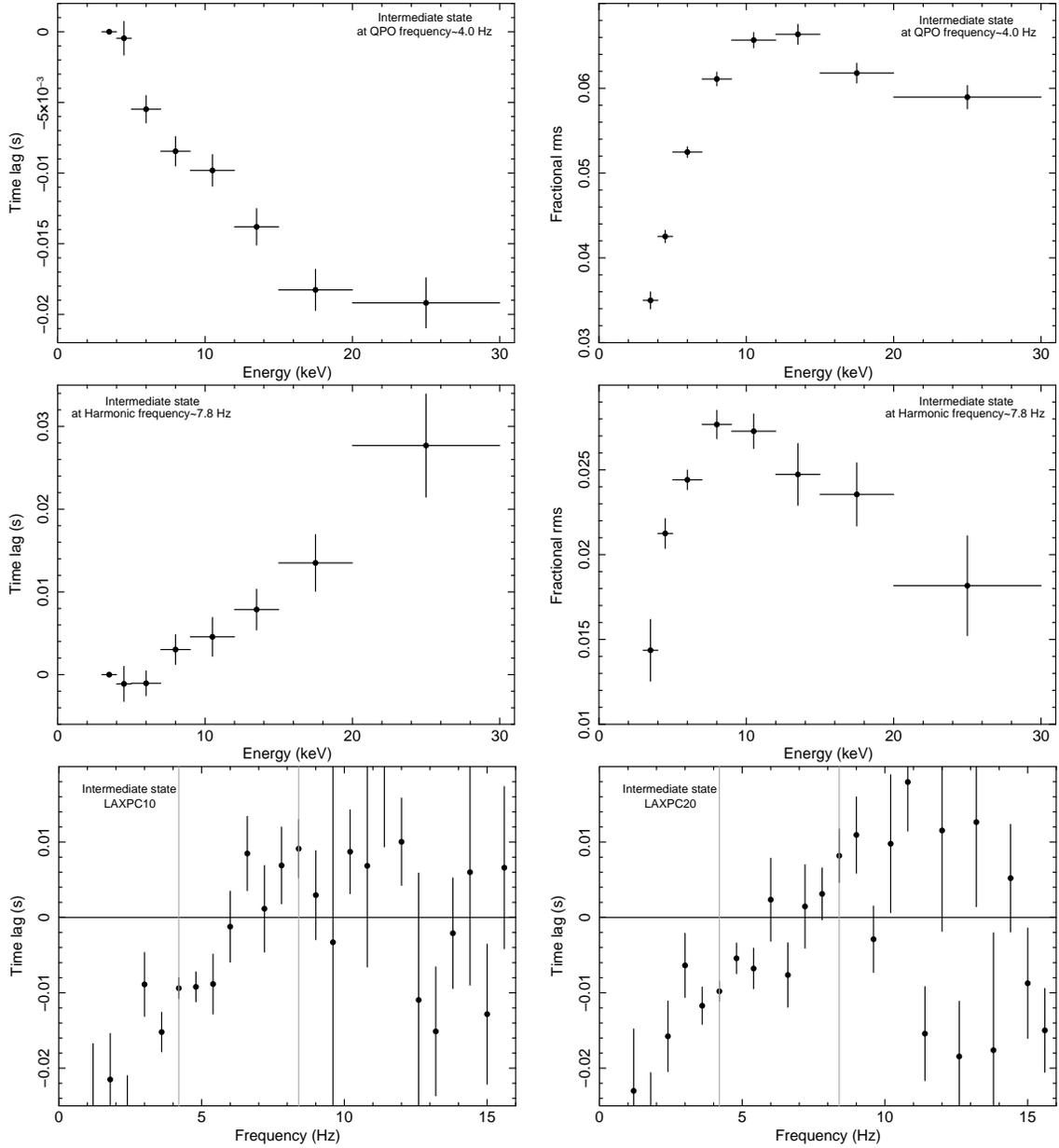
 
\centering \includegraphics[width=0.30\textwidth,angle=-90]{figure6.1.ps}
\centering \includegraphics[width=0.30\textwidth,angle=-90]{figure6.2.ps}
\centering \includegraphics[width=0.30\textwidth,angle=-90]{figure6.3.ps}
\centering \includegraphics[width=0.30\textwidth,angle=-90]{figure6.4.ps}
\centering \includegraphics[width=0.30\textwidth,angle=-90]{figure6.7.ps}
\centering \includegraphics[width=0.30\textwidth,angle=-90]{figure6.8.ps}
\caption{Time lag as a function of photon energy are shown at the QPO frequency$\sim$4.0 Hz (top left) and the harmonic frequency $\sim$7.8 Hz (middle left) during the IMS combining all LAXPC units. Fractional rms as a function of photon energy is shown at the QPO frequency$\sim$3.6 Hz (top right), the harmonic frequency $\sim$7.8 Hz (middle right) during the same combining all LAXPC units. Time lag between 3-5 keV and 9-12 keV as a function of Fourier frequency is shown in the bottom left and bottom right panels for LAXPC10 and LAXPC20 respectively. Vertical grey lines in both panels show the position of the QPO and its harmonic.}
\label{timing2}
\end{figure*}

The time lag and rms spectra shown in Figure \ref{timing2}   shows a similar trend as observed for the $\chi$ class. However, the fractional rms values at different energy bands are less than those observed during the $\chi$ class.
In bottom panels of Figure \ref{timing2}, we observe a soft lag up to 6 Hz and then the lag changes sign and becomes hard lag up to $\sim$9 Hz. This is similar to the behavior of the $\chi$ class and independently observed from LAXPC10 (bottom left panel) and LAXPC20 (bottom right panel) respectively. The dotted lines mark the QPO frequency ($\sim$4.0 Hz) and the harmonic component ($\sim$7.8 Hz) showing the soft lag and the hard lag for the QPO and harmonic component respectively.

\subsection {Heartbeat state}
\begin{figure*}
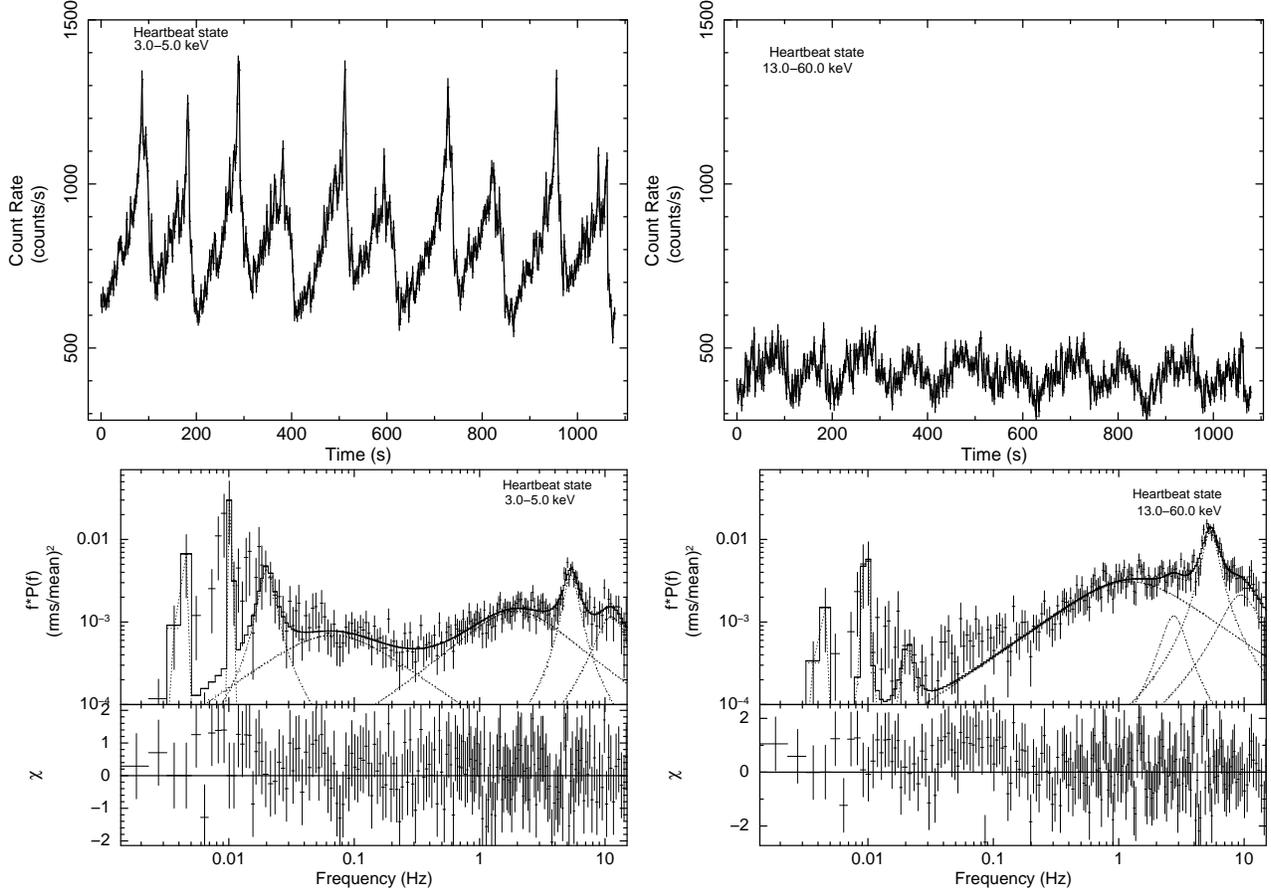

\centering \includegraphics[width=0.33\textwidth,angle=-90]{figure7.1.ps}
\centering \includegraphics[width=0.33\textwidth,angle=-90]{figure7.2.ps}
\centering \includegraphics[width=0.31\textwidth,angle=-90]{figure7.3.ps}
\centering \includegraphics[width=0.31\textwidth,angle=-90]{figure7.4.ps}
\caption{A 1100 secs lightcurve of the HS in 3.0-5.0 keV energy range when an alternate high flux, pulsed and low-flux peaked flares are observed systematically shown in top left panel. Top Right panel show the similar section but in the energy range 13-60 keV. The light curves are generated combining {\tt LAXPC10}, {\tt LAXPC20} and {\tt LAXPC30} units. Bottom panels show the power density spectra for the HS state in 3.0-5.0.0 keV  and 13.0-60.0 keV energy range fitted with multiple Lorentzians.}
\label{flaring} 
\end{figure*}

The top left panel of Figure \ref{flaring} illustrates $\sim$1100 sec lightcurve of the HS in 3.0-5.0 keV energy band where a peculiar behavior can be noticed. As the source evolves a sequence of alternately high and low peak count rate is observed where peak count rate changes by a factor of $\sim$1.5 in every 100 sec. The Lightcurve profile of the same section in 13-60 keV energy range is shown in the top right panel of Figure \ref{flaring}. The high count rate peak followed by a low count rate peak structure is absent in hard band (13-60 keV). A detailed investigation reveals that the cycle period of the flares decreases from $\sim$150 sec to $\sim$100 sec in $\sim$20 ks HS observation. Moreover, the peak count rate of flares also increases and dip count rate decreases. The bottom left and right panels of Figure \ref{flaring} show the PDS for the HS in the Fourier frequency range of 1 mHz to 2 Hz 
in the soft and hard bands respectively. Two QPOs are clearly visible: one due to the large-amplitude cyclic variation observed in lightcurve with the QPO frequency of $\sim$5 mHz and another at a frequency of $\sim$5 Hz, similar to that observed from  $\chi$ class and IMS. We see that the fractional rms for the mHz QPO is smaller in the hard band in comparison to the soft band while an opposite behaviour is seen for the LF QPO.

\begin{figure*}
\begin{center}
\centering
\includegraphics[width=0.65\textwidth,angle=-90]{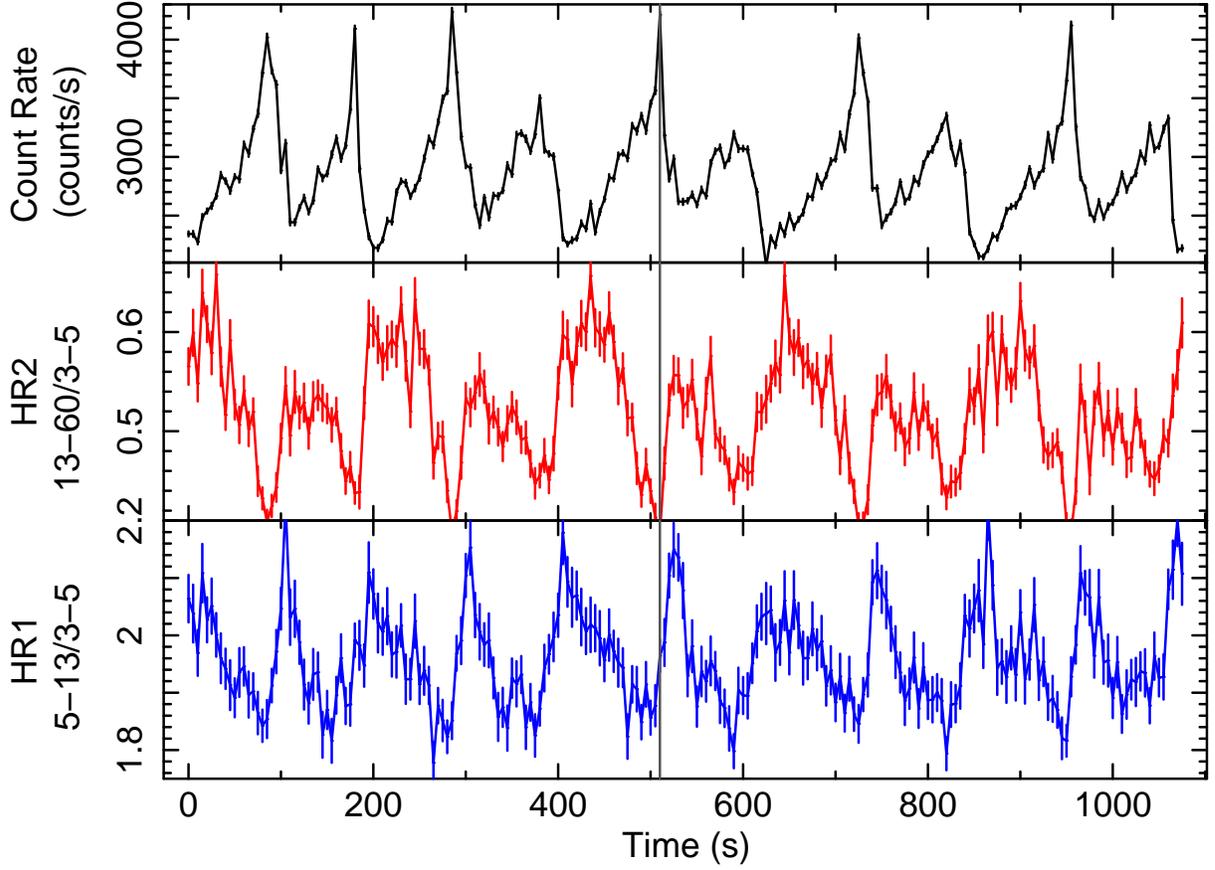}
\caption{The HS lightcurve in the 3-60 keV band with alternate large pulse and peaked flares (top panel) are shown along with hard-color (HR2) (middle panel) and soft color (HR1) (bottom panel) as a function of time using all LAXPC units. A vertical grey line is drawn along all three panels at the position of large pulse so that corresponding HR2 and HR1 can be noted. }
\label{figure8}
\end{center}
\end{figure*}

To further confirm the spectral nature of the alternative profiles in the soft band,  we plot the lightcurve, hard color (HR2) and soft color (HR1) as a function of time in Figure \ref{figure8}. A dip in HR2 is observed exactly at the time when the count rate reaches its maxima. This implies that the spike in the lightcurve is a soft pulse. A time delay of 10-15 sec is observed between the peak of HR1 and peak of lightcurve.
The implications of such results are provided in the discussion section.  
In HID plot of HS (Figure \ref{full_lc} right panel), we do not observe a loop like structure as observed by  \citet{al11} during the $\rho$ class of GRS 1915+105 and IGR J17091-3624. However, we can anticipate such behavior from Figure \ref{figure8} where the peak of the HS lightcurve profile exactly corresponds to the minimum of the hard color. The anti-correlation  between X-ray count rate and the hard color is observed as expected with zero time delay between X-ray count rate and the hard color.

\begin{figure*}
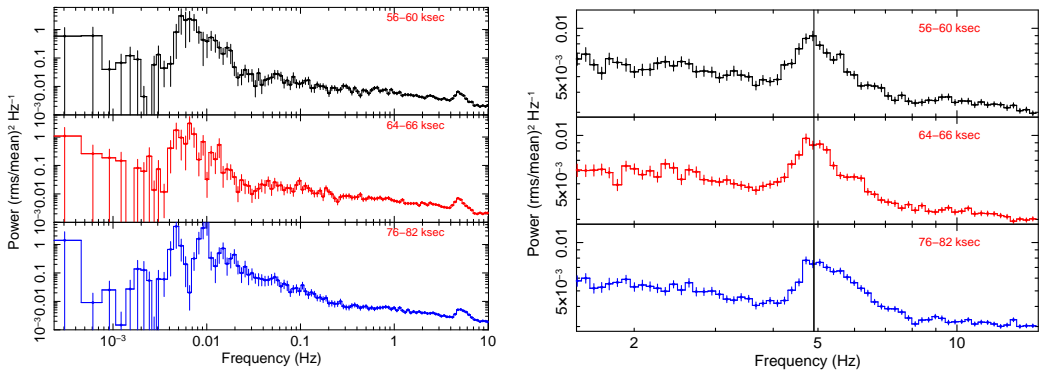

\centering \includegraphics[width=0.32\textwidth,angle=-90]{figure10.1.ps}
\centering \includegraphics[width=0.34\textwidth,angle=-90]{figure10.2.ps}
\caption{Left panel shows power density spectra of the HS in the frequency range 500 $\mu$Hz to 10 Hz (energy range 3-80 keV) during three different orbits: at the beginning of the HS cycle (top panel), at the middle of the HS (middle panel) and at the end of the HS (bottom panel). Strong QPOs due to cyclic behavior and low-frequency QPO at $\sim$6 mHz and $\sim$5 Hz can be observed clearly. Zoomed view of $\sim$5 Hz QPO during same three orbits are shown in the right panel. }
\label{p}
\end{figure*} 

The left panel of Figure \ref{p} shows PDS of the HS in the frequency range between 550 $\mu$Hz and 10 Hz. 
The PDS during the end of the HS observation (76-82 ks in Figure \ref{lightcurve}) shows the QPO peak and its harmonic. A zoomed view (right 
panel of Figure \ref{p}) of low-frequency QPO $\sim$5 Hz in three different times of observation shows that the peak of the QPO frequency does not vary within the HS. However, the shape changes. During the last orbit of  HS observation (76-82 ks),  we notice a broad, distorted and  asymmetric profile for LFQPO while two peaks in the profile structure of mHz QPO becomes more prominent.
However, it is not clear whether the distortion and broadening in the QPO profile are
because of change in QPO frequency with time or due to the presence of two QPO's separated by a small frequency interval which we cannot resolve.

\begin{figure*}
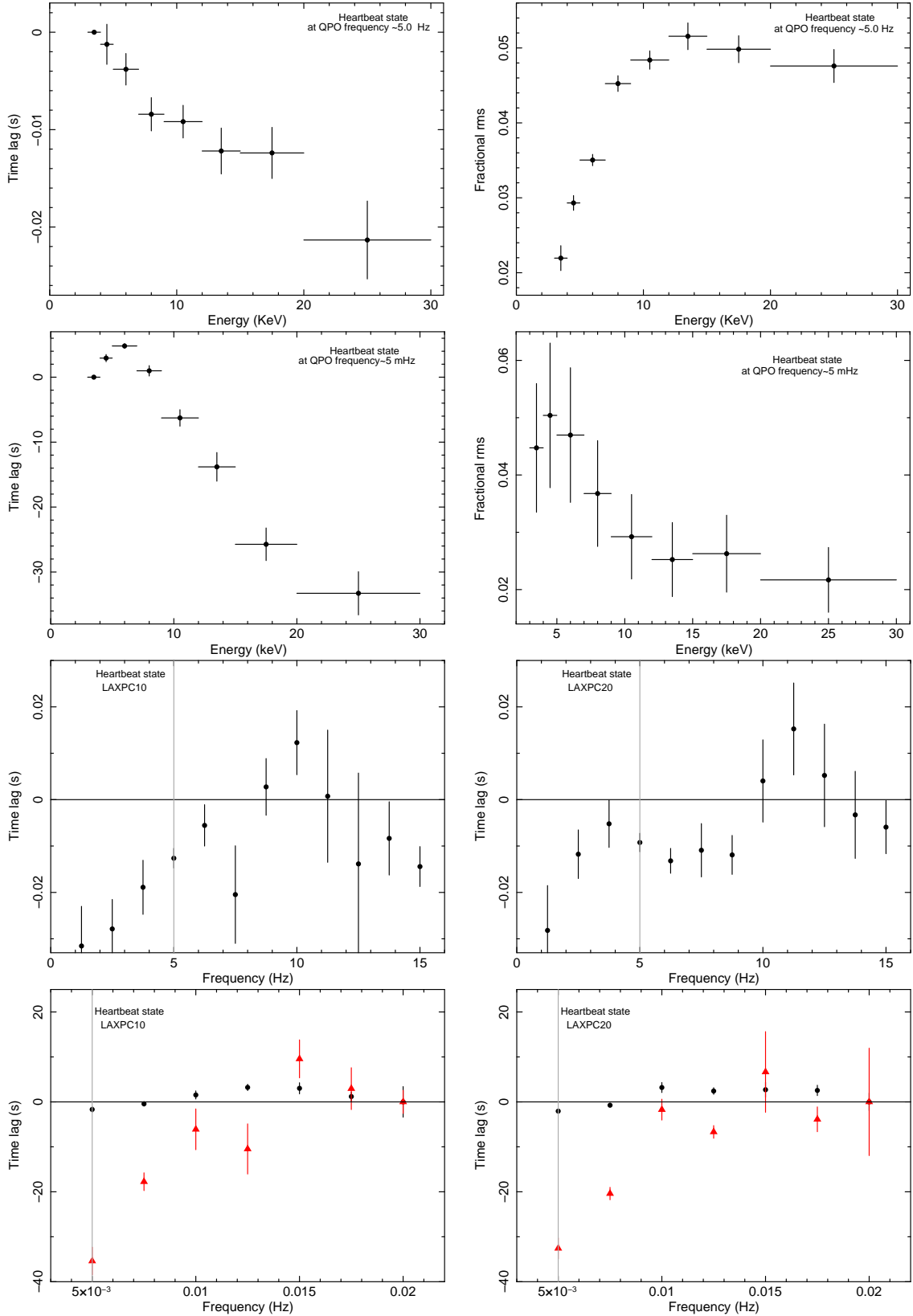

\centering \includegraphics[width=0.30\textwidth,angle=-90]{figure11.1.ps}
\centering \includegraphics[width=0.30\textwidth,angle=-90]{figure11.2.ps}
\centering \includegraphics[width=0.30\textwidth,angle=-90]{figure11.3.ps}
\centering \includegraphics[width=0.30\textwidth,angle=-90]{figure11.4.ps}
\centering \includegraphics[width=0.30\textwidth,angle=-90]{figure11.5.ps}
\centering \includegraphics[width=0.30\textwidth,angle=-90]{figure11.6.ps}
\centering \includegraphics[width=0.30\textwidth,angle=-90]{figure11.7.ps}
\centering \includegraphics[width=0.30\textwidth,angle=-90]{figure11.8.ps}
\caption{Top left and top right panels show time-lag and fractional rms as a function of photon energy at the LF QPO frequency $\sim$5 Hz during the HS using all LAXPC units. Upper middle left and upper middle right panels show the energy-dependent time lag and frectional rms at mHz QPO frequency. Time lag of 9.0-12.0 keV photons with respect to 3.0-5.0 keV photons as a function of frequency is plotted for LAXPC10 and LAXPC20 in lower middle left and lower middle right panels respectively. The lag at the QPO frequency of $\sim$5 Hz is shown by a vertical grey line. Bottom panels shows time lag as a function of frequency for the energy 
band 5.0-13.0 keV 
relative to the band 3.0-5.0 keV (black coloured data points), 13.0-60.0 keV 
band relative to 3.0-5.0 keV band (red color data points) and 13.0-60.0 keV band
relative to 5.0-13.0 keV band (blue color data points).}
\label{fslag}
\end{figure*} 

Time-lag as a function of photon energy for the QPO at $\sim$5.0 Hz is plotted in the left panel of Figure \ref{fslag} while the right panel shows the fractional rms spectra for the same QPO.  Similar to $\chi$ class and IMS, the lag spectra show soft lag while fractional rms increases with energy. The energy-dependent time lag spectra and the rms spectra at the Fourier frequency of ~5 mHz are shown in the upper middle left and upper middle right panels of Figure \ref{fslag}. The energy-dependent lag spectra show hard lag up to $\sim$6 keV and soft lag at higher energy.  However, the rms spectra show opposite behaviour. During the HS, the fractional rms strictly decreases with energy while the energy-dependent fractional rms during the $\rho$ class either increases with energy or initially increases up to $\sim$10 keV and then decreases \citep{mi16}. The time lag versus frequency plot shows the same behaviour at QPO ($\sim$5.0 Hz) and harmonic frequency as observed in  $\chi$ class and IMS. The bottom left and bottom right panels of Figure \ref{fslag} shows time lag of 5.0-13.0 keV energy band with
respect to 3.0-5.0 keV band (black data points), 13.0-60.0 keV band with
respect to 3.0-5.0 keV band (red data points) and 13.0-60.0 with respect to 5.0-13.0 keV photons (blue data points) as a function of Fourier frequency for LAXPC10 and LAXPC20 respectively at QPO frequency $\sim$5 mHz. For 13.0-60.0 keV band, a soft lag is observed with respect to both 3.0-5.0 keV and 5.0-13.0 keV bands for mHz QPO. In contrast, the lag is very small for 5.0-13.0 keV relative 3.0-5.0 keV band at mHz QPO frequency. 
\subsection{Dynamic Power Spectra}
\begin{figure*}
\centering \includegraphics[width=0.22\textwidth,angle=-90]{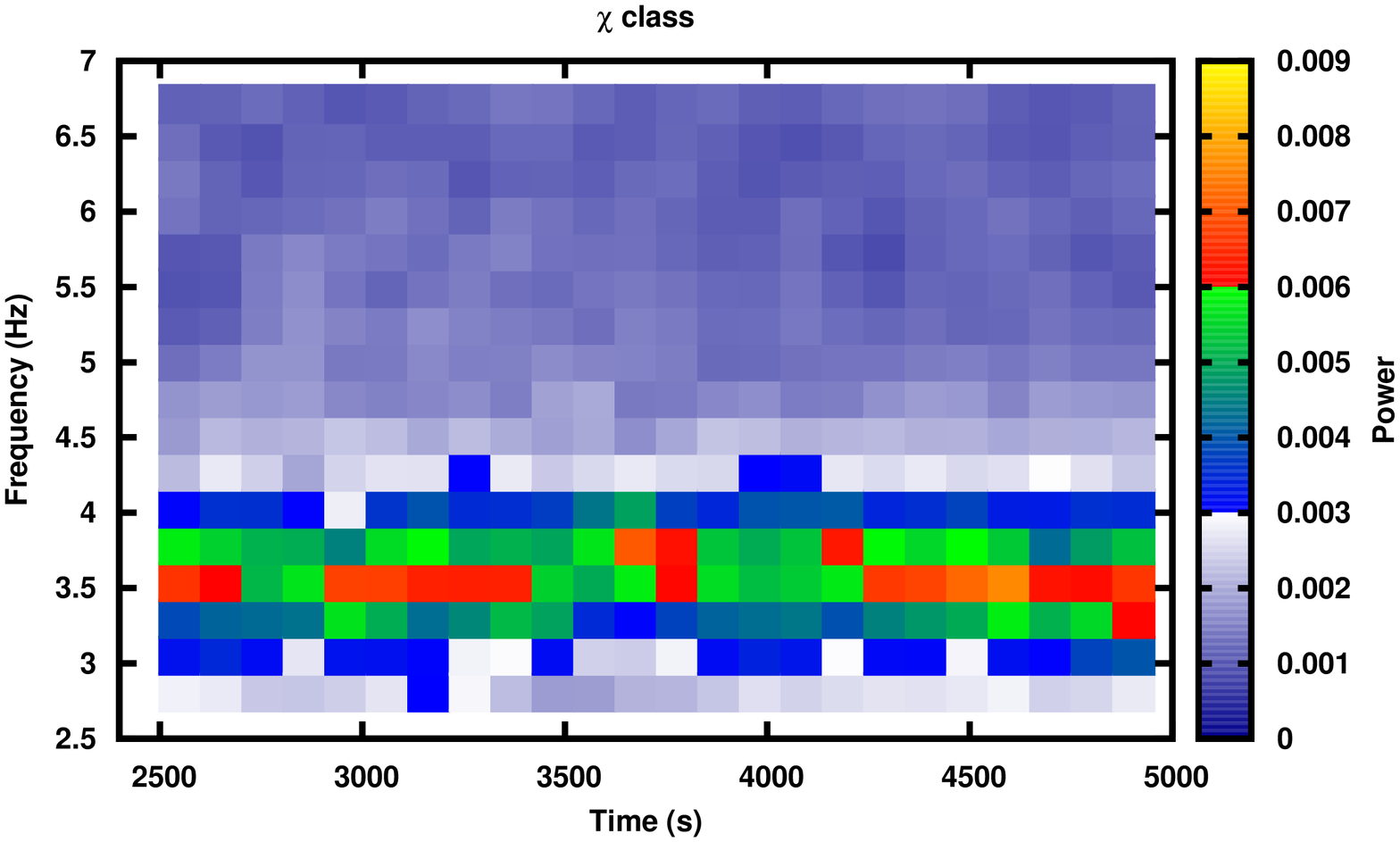}
\centering \includegraphics[width=0.22\textwidth,angle=-90]{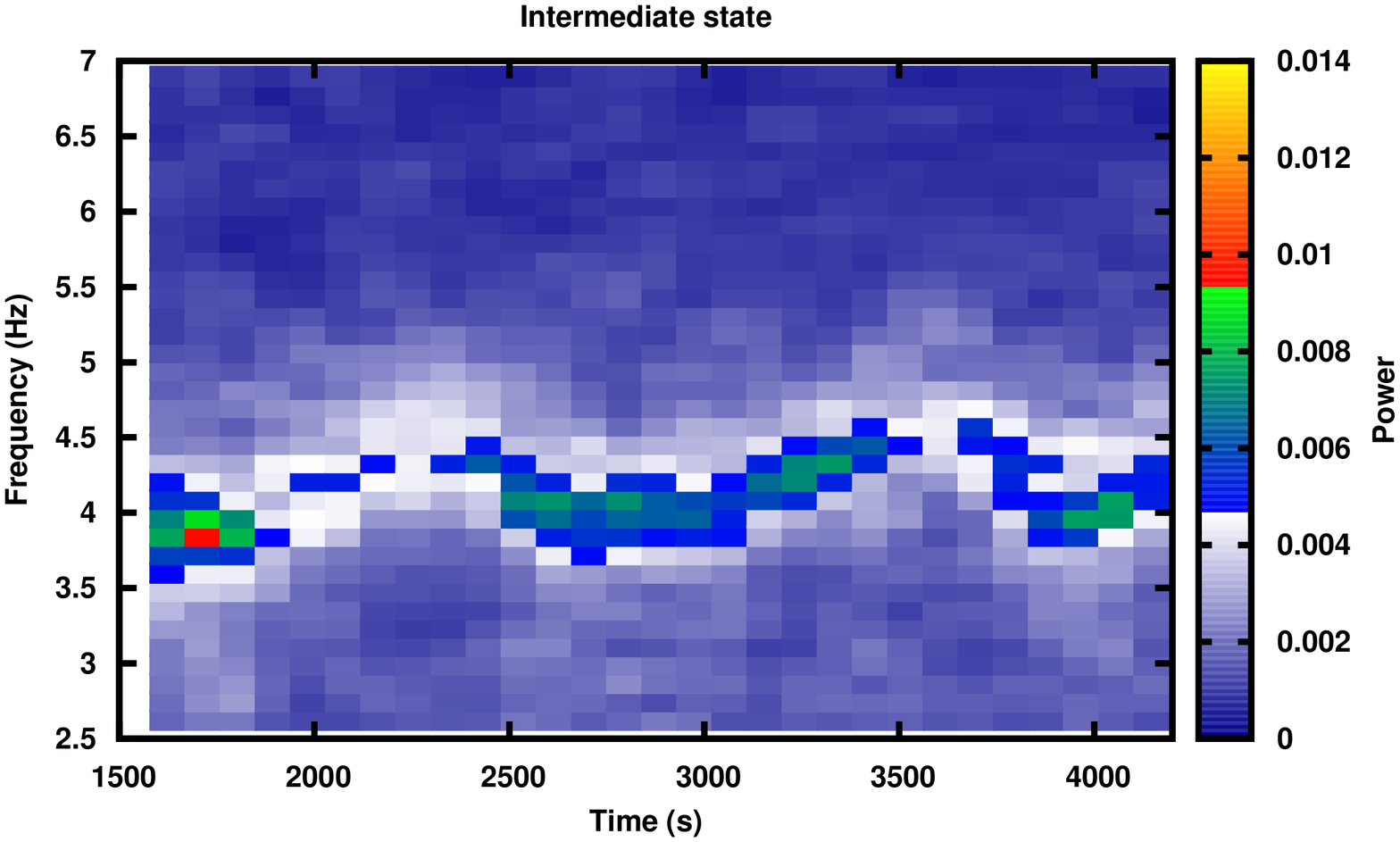}
\centering \includegraphics[width=0.22\textwidth,angle=-90]{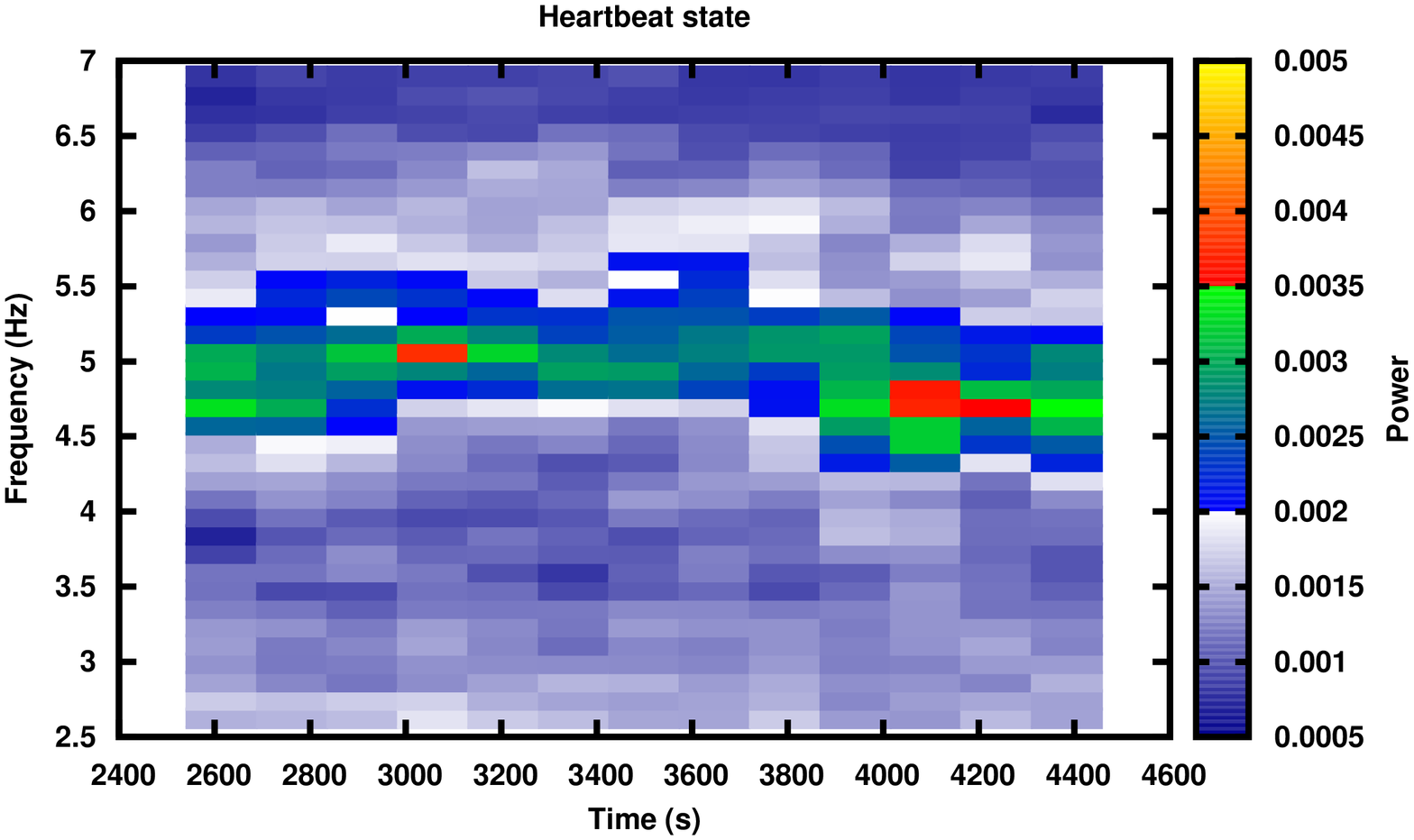}
\caption{3-30 keV Dynamic Power Spectra during the $\chi$ class (left panel), IMS (middle panel) and HS(right panel) are shown in the frequency range 2.5-7 Hz. A gradual increase in the QPO frequency and its time-dependent behavior is clearly observed from three panels.  }
\label{dynamo}
\end{figure*}

In order to check the time-dependent behavior of the QPOs in different states, we computed the dynamic power spectra (DPS). We plot the DPS for $\chi$ class  (left panel), IMS (middle panel) and HS (right panel) in the energy range 3-80 keV and frequency interval 2.5-7.0 Hz in figure \ref{dynamo}. During $\chi$ class , we observe the peak of the QPO frequency at $\sim$3.5 Hz nearly constant in time. However, the rms power of QPO show somes variation with time. In contrast, a systematic variation in the QPO frequency between 3.5 Hz and 4.5 Hz has been observed in the DPS of IMS. Moreover, we also see systematic variations in the strength of the QPO. When the QPO frequency shifts to the highest value the rms power decreases while the rms power is highest close to the minimum of the QPO frequency. During both cycles, each having a period of $\sim$1500 sec, the rms power changes by a factor of $\sim$2. Double cyclic change is observed in both QPO frequency and the rms of the QPO in an anti-correlated manner. However, we do not see any cyclic pattern in the DPS when the source transits from the IMS to HS. As observed from the plot, the QPO frequency changes between 5.5 Hz and 4.5 Hz in 2000 sec, however, the rms does not change significantly in this timescale. Interestingly, the peak rms power of the HS is smaller by a factor of $\sim$2 when compared to that of $\chi$ class and IMS. 
 
\section{Discussions}              
In this work, we present the continuous monitoring of GRS 1915+105 ($\sim$ 90 ks) using \asat{}/LAXPC when a weak to strong variability transition is observed in X-rays. A hardness-intensity diagram (HID), plotted for the entire observation shows a gradual transition from the spectrally hard to a spectrally softer state (Figure \ref{full_lc}). Three different states in the color-color diagram are distinguishable from each other (right panel of Figure \ref{full_lc}). During the first few tens of ks, the source shows no large-amplitude, long timescale (few tens of seconds) variabilities in both soft (3-5 keV) and hard (13-60 keV) X-ray light curves. This variability is similar to the radio-quiet $\chi$ class ($\chi_{RQ}$ ). The power density spectrum in soft band reveals the presence of a strong low-frequency QPO at $\sim$3.6 Hz and its harmonic at $\sim$7.2 Hz. However, in the hard band the harmonic component is not significant. We computed the time-lag spectra and the fractional rms spectra as a function of photon energy at the QPO and harmonic frequencies. The time delay spectra show the soft lag trend at the QPO while at the harmonic frequency we observe the hard lag.

\subsection{Comparison between observed $\chi$ class and other subclasses of the $\chi$ class}
Using a radio/X-ray joint monitoring program, \citet{ro08b} observed that the 15 GHz radio flux density during $\chi$ X-ray class gradually changed from a radio-quiet state with a mean level of $\sim$44.9 mJy to a radio-loud state with a mean level of $\sim$70.4 mJy. Later \citet{pa13} showed that the radio flux density during the radio quiet $\chi$ class can be as low as 10 mJy or less. Although the classification is based on the radio flux density, the X-ray spectral and timing properties of the radio-quiet and the radio-loud $\chi$ class are remarkably different. Although no simultaneous radio observation was performed with the AstroSat observation of the $\chi$ class presented here, we observed that the X-ray timing properties, e.g., QPO frequencies and strength, the nature of lightcurve, CCD, HID, lag spectra of the $\chi$ class observation analyzed in this work is remarkably similar to the radio-quiet $\chi$ class and dissimilar to the radio-loud $\chi$ class. We discuss them below.
During the radio-loud $\chi$ class (plateau state) no phase reversal is observed between the QPO and the harmonic frequencies in GRS 1915+105 when QPOs are observed at a frequency lower than 2 Hz \citep{tr01, ya06, pa13}. On the other hand, during the radio-quiet $\chi$ class as observed by \citet{pa13} phase reversal is noticed during the QPO and the harmonic frequencies and QPOs are observed at a frequency similar to that observed during our present $\chi$ class observations. We also notice phase reversal in the lag spectra.  
However, there are differences between the $\chi$ class observed here and other subclasses of the $\chi$ class. During the $\chi$ class, we observe an additional strong low-frequency broad Lorentzian noise component at a frequency of $\sim$0.25 Hz which is absent from the PDS of the  radio-quiet $\chi$ class in GRS 1915+105 observed with \asat{}/LAXPC \citep{ya16b} and \xte{}/PCA \citep{pa13}. Secondly, for the QPO component, the time lag spectra of the $\chi$ class show a soft lag accompanied by sharp upturn at 20 keV above which it shows hard lag \citep{ya16b}. In the present analysis, we do not observe any hint of an upturn in the time lag spectra for the QPO component at higher energies in the $\chi$ class. Owing to its high efficiency, LAXPC is able to resolve more complex structure in the PDS at different photon energies and able to constrain lag spectra at harder X-ray band than \xte{}/PCA can do. Therefore, such differences can be attributed to the high detection efficiency of LAXPC compared to \xte{}/PCA, particularly in the harder band. It may be noted that during the $\chi$ class, the 3.0-80.0 keV X-ray intensity using {\tt LAXPC10} is $\sim$860 counts/s while it was much higher($>$2000 counts/s) in earlier observed radio-quiet $\chi$ class using \asat{}/LAXPC \citep{ya16b}. 

\subsection{Transition phase}

At the end of the $\chi$ class, marked by the sudden increase and strong variability in the hardness (bottom panel of Figure \ref{lightcurve}), we note that a long-term, large amplitude, irregular variability is introduced in the lightcurve (top left panel of Figure \ref{intermediate}). Such variability on the time scale of a few thousands of seconds is only observed in the soft X-ray band (3-5 keV) but absent from the hard X-ray band lightcurve (13-60 keV). We termed it as `Intermediate state (IMS).' The large-amplitude, long-timescale, irregular variability in the soft X-rays is possibly linked to the onset of an outer disk ($>$50 R$_g$) turbulence caused by the fluctuations in the mass accretion rate \citep{ne11}. Despite large amplitude fluctuations, it is interesting to note that the PDS of IMS shows the low-frequency QPO at $\sim$4 Hz and it's harmonic while the QPO frequency is slighter higher than that observed during the $\chi$ class. The strength of the QPO and its harmonic during both IMS and $\chi$ class are similar. Moreover, the time-lag spectra of IMS and $\chi$ class are also very similar to each other. The reduced fractional rms values in IMS in comparison to those observed during the $\chi$ class may imply that with the launch of large-amplitude, irregular flares, the strength of the QPO is reduced.   Comparing hard X-ray lightcurve of $\chi$ class and IMS, we may note that the hard X-ray HR2 remains the same for both states where HR1 attains the highest value in IMS.
From the DPS, we may note that during IMS, the LF QPO surprisingly disappears during certain time intervals. The time-averaged QPO frequency, rms, and width, of the observed LF QPOs are similar to type-C QPO. However, the type-C QPO are found to be very stable for other X-ray binaries in contrast to our observations. We speculate that the quasi-appearance and disappearance of type-C QPO during IMS may be caused by the inner disk perturbation driven by the large amplitude, slow outer disk variability. 

\subsection{Heartbeat state - Similarities and dissimilarities with classical $\rho$ class}

Immediately after the IMS, semi-regular periodic fluctuations have been observed during the last phase ($\sim$20 ks) of our observations.  We termed it as the 'Heartbeat state (HS)' for the following reasons. The variability profile is highly structured and has a significant resemblance to the `$\rho$' class/heartbeat oscillations in GRS 1915+105 \citep{ya99,be00,ne11}. For example, during the HS in our observation, the ratio of maximum-to-minimum flux, averaged over ten cycles, increases from $\sim$1.5 to $\sim$2.1 with time while the same during a typical $\rho$ class changes from $\sim$1.73 to $\sim$3.33 with time. Although the trend is the same, values are different. As these ratios are instrument-independent, lower ratio during our observation implies that the lightcurve variability is not fully evolved to a classical $\rho$ class. A similar trend but difference in the evolution of cycle timescale values are observed between the HS and typical $\rho$ class. During HS, the cycle timescale ($\sim$150 sec) in first few ks is found to be longer than that during last few ks of observations ($\sim$100 sec) which is shown in the top panel of Figure \ref{figure8}. An evolution of $\rho$ cycles towards faster timescale is also observed from GRS 1915+105 (110-40 sec) and another black hole X-ray transient IGR J17091-3624 (50-20 sec) \citep{pa14}. However, in both cases, the evolution started from large cycle period and evolved up to about half of its value (Figure 6  and Table 3 in \citet{pa14}). Such offset in values with similar trends implies that given a sufficient time to evolve, the HS state variability will eventually lead to the $\rho$ class. It may also be noted that each flare in $\rho$ class shows two peaks when its time cycle was fast (like  for ~50 s cycle time for GRS 1915+105) (Figure 7 in \citet{pa14}).
In our observation of HS,  it is usually single peak with time cycle $>$100s. We also observe a variation in the alternate peak count rate as shown in top left panel of Figure \ref{flaring} . Such a
behavior is seen for the first time in GRS
1915+105. Therefore, we observe here an evolution of the large-amplitude, systematic variability which leads to the onset of the heartbeat oscillations in GRS 1915+105. Hence the HS stage can be considered as the precursor of the heartbeat oscillations. We observe a steady increase in lightcurve fractional rms from $\chi$ class to HS as shown in Table \ref{obs1}. This can be attributed to the long term-variability in the light curve.

Despite few similarities in the evolution trend, few interesting differences between the HS and the $\rho$ class/heartbeat oscillations are noteworthy. We observe single-peaked pulse during the HS. However, the flux during the pulse peak changes by a factor of $\sim$1.4 in alternate peaks shown in the top panel of  Figure \ref{figure8}. Such a large and systematic change in flux in alternative peaks has not been reported for typical $\rho$ class observations. The nature of the hard color (HR2) and the soft color (HR1) during single-peaked pulse in $\rho$ class is shown by \citet{ne12}. At the position of the pulse peak, shown in their Figure 1, HR2 shows a sudden dip while HR1 shows a sharp peak. This is expected since the pulse is soft. While the position of the HR2 dip of the pulse during our HS observation in the middle panel of  Figure \ref{figure8} matches with that from \citet{ne12}, the position of HR1 shows a clear discrepancy by ∼15-20 sec. The energy-dependent lag spectra at the mHz QPO frequency shown in the upper middle left panel of Figure \ref{fslag} also exhibits a peculiar trend of hard lag of the order of few seconds up to ~6 keV and then a soft lag of the order of few tens of seconds at harder X-ray bands. Similar behaviour has also been observed from the RXTE/PCA time lag spectra of $\rho$ class in GRS 1915+105 at the heartbeat frequency \citep{mi16}. The rms spectra for the HS show
different behaviour in comparison to that observed by \citep{mi16} for $\rho$ class. The reason for such discrepancy is not clear. The lag timescale is so large that it can be accounted neither by the reprocessing nor by the Comptonization timescale. Therefore, propagation of fluctuations in viscous timescale \citep{ut11} and a delayed response of the inner disk radius in response to fluctuating mass accretion rate as observed by \citet{mi16} are suitable for explaining the long lag.
 
\subsection{QPO evolution in three states}      

Although the long-term, large-amplitude component with the variability timescale of the order of few hundreds of seconds significantly evolves from the $\chi$ class to HS, the time-averaged low-frequency QPO properties  change marginally as seen in Tables \ref{obs2}, \ref{obs3} and \ref{obs4}. 
In the 3-5 keV energy band, we observe that from the $\chi$ class to HS, the low-frequency QPO changes its frequency from $\sim$3.6 Hz to $\sim$5.3 Hz, while its fractional rms and the width, averaged over few ks, changes from $\sim$5.7\% to $\sim$4.3\% and $\sim$0.56 Hz to $\sim$1.56 Hz respectively (Table \ref{obs2}). Similar changes are seen in other energy bands. The time-lag spectra show soft lag during all three states, and the rms spectra are very similar. 
Therefore, if we assume that the low-frequency QPO is caused by the inner disk activity in a coupled disk-corona system, then the large-amplitude variability may be an outer disk phenomena and the long timescale variability does not perturb short timescale variability originated from the inner accretion disk. Such a scenario is also consistent with \citet{ne11} who assume that the fluctuation during $\rho$ cycles which causes the density wave that propagates inwards, originates at a disk radius of 30 R$_g$ or higher. Interestingly, the harmonic component which is strongly detected during the $\chi$ class and IMS becomes relatively insignificant during the HS, as can be seen from figure \ref{flaring}. We also see from Tables \ref{obs2} and \ref{obs3} that the width of the harmonic becomes large in the HS which
would lead to a relatively small Q-factor. 
During the $\chi$ class and the IMS, 
the time-lag and the rms spectra at the harmonic frequency 
are found to be similar. 

\section{Summary and Conclusions}
\begin{itemize}
\item {In this paper, we present AstroSat observation of a peculiar galactic microquasar GRS 1915+105 and on the basis of  change in the variability of  X-ray light curve, CCD and HID plots we report a state transition from $\chi$ class to a heartbeat state via an intermediate state.}
\item{The higher sensitivity and resolution  of AstroSat/LAXPC as compared to RXTE/PCA data unveil some new features of $\chi$ class. Therefore, with LAXPC we note that the subclasses of the $\chi$ class behave differently at different X-ray intensity level. }

\item {The consistency of PDS, time-lag, rms spectral properties and the hard X-ray flux between $\chi$ class and IMS imply that the large-variability observed in soft band IMS does not affect the hard X-ray variability. The low-frequency QPOs, corresponding harmonics as well as nature of lag and rms spectra which are inner accretion properties remain largely unaffected by such outer disk amplitude modulation.}

\item { An interesting feature that we observe is the transient nature of the C-type QPO during the IMS phase. 
We find that the dynamic power spectra of IMS in the energy range 3.0-30.0 keV (middle panel of Figure \ref{dynamo}) show a remarkable evolution of the QPO strength by a factor of $\sim$2 which is probably driven by the long-term disk fluctuation. We suggest that this may be associated with the inner disk perturbation caused by large amplitude variation of the outer disk. }
\item{
A preliminary time-averaged spectral analysis of the HS, which would be discussed elsewhere, indicates that the Eddington luminosity fraction, defined as the ratio between 0.1-100 keV unabsorbed X-ray luminosity to the Eddington luminosity, is $\sim$0.2-0.25 while the same during typical $\rho$ class is $\sim$0.7-0.8 \citep{ne11,ne12}. This indicates that the inner accretion region in typical $\rho$ class is strongly dominated by the radiation pressure compared to the inner disk during HS. Therefore, the difference in the soft/hard lag behaviour between the HS and the $\rho$ class may be attributed to the modifications of the accretion and radiation geometry caused by the strong radiation pressure at the inner disk region. Exploring further in this direction is beyond the scope of the present work. }
\end{itemize}

\section{Acknowledgment}
We thank the referee for constructive suggestions which were very helpful in improving the paper. We thank  members of LAXPC instrument team for their contribution to the development of the LAXPC instrument. We also acknowledge contributions of the AstroSat project team at ISAC. This paper makes use of data from the AstroSat mission of the Indian Space Research Organisation (ISRO), archived at the Indian Space Science Data Centre (ISSDC). MP acknowledges TIFR for giving him three months visiting position and the Royal Society-SERB Newton International Fellowship support funded jointly by the Royal Society, UK and the Science and Engineering Board of India(SERB) through Newton-Bhabha Fund.

\end{document}